\documentclass[]{rsos}
\usepackage{graphicx}
\usepackage{epstopdf, epsfig}
\usepackage{amsmath}
\usepackage{subfigure}

\begin{document}

\title{Quasi-two-layer morphodynamic model for bedload-dominated problems: bed slope-induced morphological diffusion}

\author{Sergio Maldonado$^{1}$ and Alistair G. L. Borthwick$^{2}$}
 
 \address{$^{1}$Faculty of Engineering and the Environment, University of Southampton, Highfield, Southampton SO17 1BJ, U.K.\\
 	$^{2}$Institute for Energy Systems, School of Engineering, University of Edinburgh, The King's Buildings, Edinburgh EH9 3JL, U.K.}
 
 \subject{Civil Engineering, Mathematical modelling, Geophysics}
 
 
 \corres{Sergio Maldonado\\
 	\email{s.maldonado@soton.ac.uk}}
 
\begin{abstract}
	
	We derive a two-layer depth-averaged model of sediment transport and morphological evolution for application to bedload-dominated problems. 
	The near bed transport region is represented by the lower (bedload) layer which has an arbitrarily constant, vanishing thickness (of approximately ten times the sediment particle diameter), and whose average sediment concentration is free to vary. 
	Sediment is allowed to enter the upper layer, and so total load may also be simulated, provided that concentrations of suspended sediment remain low. 
	The model conforms with established theories of bedload, and is validated satisfactorily against empirical expressions for sediment transport rates and the morphodynamic experiment of a migrating mining pit by \cite{Lee1993}. 
	Investigation into the effect of a local bed gradient on bedload leads to derivation of an analytical, physically meaningful expression for morphological diffusion induced by a non-zero local bed slope. 
	Incorporation of the proposed morphological diffusion into a conventional morphodynamic model (defined as a coupling between the shallow water equations, Exner equation and an empirical formula for bedload) improves model predictions when applied to the evolution of a mining pit, without the need either to resort to special numerical treatment of the equations or to use additional tuning parameters.
	
\end{abstract}


\begin{fmtext} 

\end{fmtext}


\maketitle

\section{Introduction} \label{sec:intro}

The majority of coastal and river morphodynamic models employed in practice consists of coupled systems of depth-averaged flow mass and momentum equations, a bed-update equation, and a sediment-transport formula. 
We refer to this type of model as a Conventional Morphodynamic (CM) model. 
A feature common to almost all CM models is the use of empirical formulae for estimation of sediment transport rates, necessary for closure of the morphodynamic model. 
However, the vast number of such formulae available in the literature makes selection of the most appropriate expression difficult, leading to considerable uncertainty. 
For example, sediment transport in open channel flows is typically divided into bedload and suspended load; but, although different mechanisms govern these two modes of transport, a reliable method to distinguish one from the other has yet to be provided \cite{Amoudry2011}. 
Therefore, ambiguity in identification of the mode of transport present (i.e. bedload vs suspended load) is reflected in the selection of the closure formula, thus adding to uncertainty. 
Further uncertainty in CM models also arises from our lack of understanding of the fundamental mechanics behind sediment transport. 
In particular, the effect of bed slope on sediment transport should  be included in reliable morphodynamic models of any kind \cite{Chiodi2014,Moulton2014}. 
However, the bed slope effect is often neglected, or accounted for through additional tuning parameters, adding empiricism to CM models and thus uncertainty when no data are available for calibration.

Alternative morphodynamic models include two-phase and two-layer models, which may be scientifically more insightful than CM models, but at the cost of increased mathematical complexity and computational demand. 
Thus, such alternatives tend to be more appealing to a scientific audience than to the practitioner community. 
Among 2-layer models, it is worth noting the early work by \cite{Fraccarollo2002}, who introduced the idea of an erosion rate estimated using simple concepts from open channel hydraulics and soil mechanics, replacing part of the empiricism inherent to sediment transport formulae with physical mechanisms that drive bed erosion. 
The model comprised clear water flowing on top of a constant-density sediment-water-mixture, which in turn had the same average density as the non-mobile bed underneath; both fluid layers moved at the same speed. 
Later, \cite{Spinewine2005} extended the work by \cite{Fraccarollo2002} to account for the different velocities and concentrations in the two fluid layers, while assuming that the average density of the transport layer was constant. 
The latter restriction was removed by \cite{Li2013}, who considered a variable-density lower layer; the variability in density was in turn estimated via an empirical formula for sediment transport. 
All the aforementioned models simulate clear water over a transport layer, and track the evolution of a distinct physical interface dividing both layers. 
(For a comprehensive review of depth-integrated two-layer models see \cite{Iverson2015}.) 
Two-phase models are also strong candidates for simulating morphological evolution and sediment transport, yielding interesting insights \cite{Bakhtyar2009,Greco2013,Chiodi2014}. 
However, we restrict our present study to depth-averaged two-layer approaches. 

In an attempt to reconcile the scientific and practitioner communities interested in morphological modelling, we propose a simplified 2-layer morphodynamic model, which may also be used to enhance CM approaches. 
The model idealises shallow water-sediment flow as being divided into two layers with temporally and spatially varying densities in the plane parallel to the mean bed. 
In order to deal with the inherent ambiguity in the distinction between different modes of transport, the thickness of the lower layer is fixed at a small value, distinguishing the present model from previous two-layer approaches. 
Then, in the spirit of \cite{Fraccarollo2002}, simple constitutive equations are used to estimate the driving mechanism of bed erosion and other closure terms, such that selection of a particular sediment transport formula is not required. 
The model is primarily designed for bedload-dominated sedimentary processes, although it is demonstrated that suspended load can also be simulated provided the suspended sediment concentration is low. 
The model is compared against an experimental study of bedload-driven migration of a mining pit, and then employed to derive an analytical, phenomenologically meaningful expression (free from tuning parameters) for the  bed slope effect on bedload. 
This expression can be included in CM models, enhancing their accuracy without increased empiricism, as verified after comparison against the migrating pit experiment.

The body of the paper is organised in three parts. 
The first part (\S \ref{sec:model-description}) deals with the description of the proposed model and its underlying assumptions. 
Analysis, as well as experimental and mathematical validation of the model are included in the second part (\S \ref{sec:model_validation}). 
The third part (\S \ref{sec:bedslope}) deals with the effect of bed slope on bedload, presenting the derivation and application of a slope-related term that can be employed within CM models; comparison against morphodynamic experimental data is also presented. 
\S \ref{sec:conclusions} summarises the key findings.

\section{The model} \label{sec:model-description}

\subsection{Description} \label{sec:phen-assumptions}

\begin{figure}
	\centering
	\includegraphics[width=0.8\linewidth]{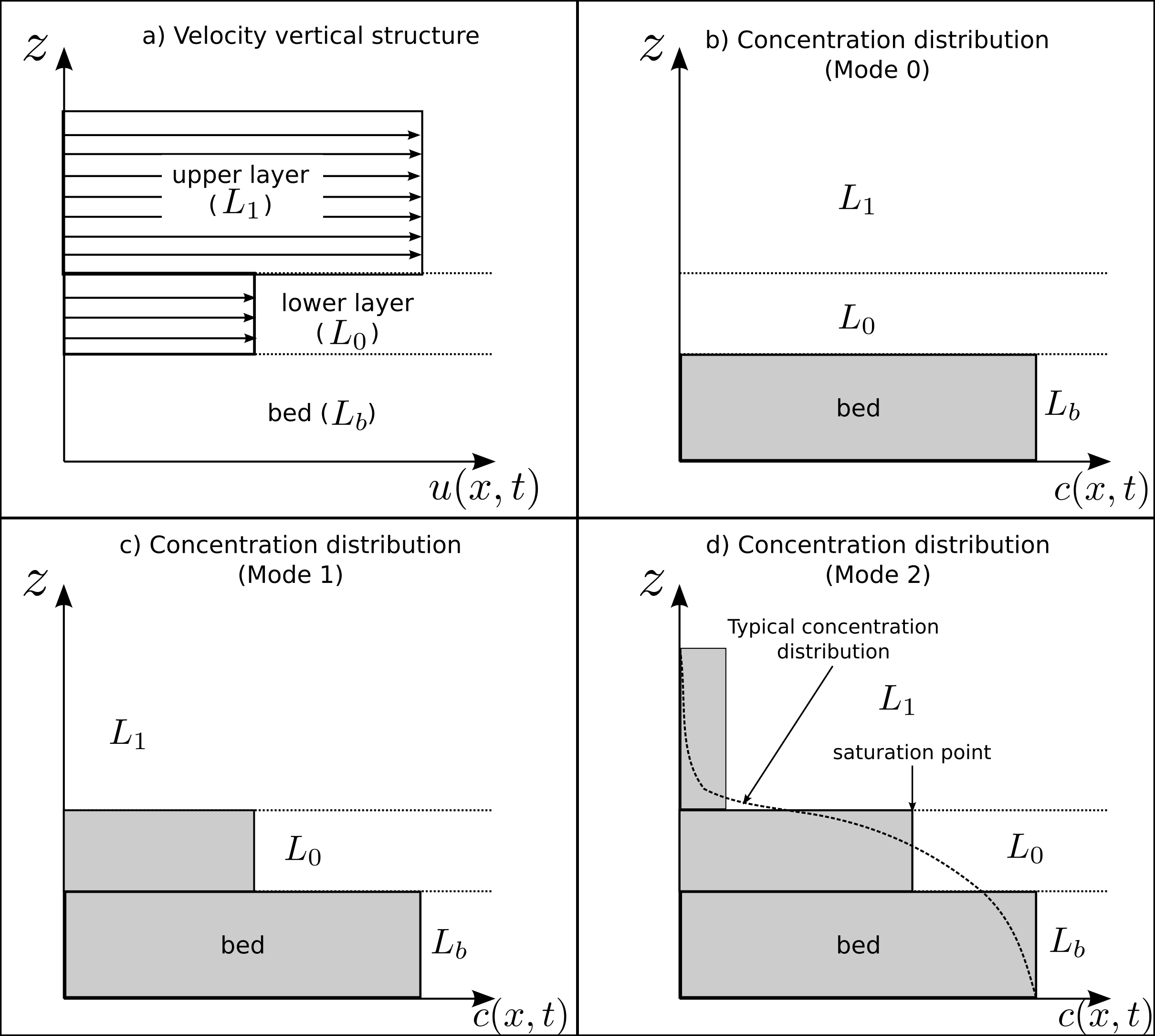}
	\caption{Assumed vertical structures for velocity and concentration: a) vertical profile of fluid horizontal velocity; b), c) and d), vertical profile of sediment concentration for transport Modes 0, 1 and 2.} \label{fig_vertical_structure}
\end{figure}

A Cartesian frame of reference $(x,z)$ is adopted, where $x$ and $z$ are the streamwise and vertical coordinates, respectively (the transverse coordinate is not considered in this paper); and time is denoted by $t$. 
The water-sediment mixture is divided into two layers: the lower concerned with bedload transport, and the upper representing sediment in suspension. 
Although the model is primarily designed to deal with bedload transport, relaxation of the assumption --- common to most 2-layer models --- of an upper water layer, allows suspended load to be considered at dilute concentrations. 
The water-sediment mixture is assumed to be an incompressible continuum, with each layer experiencing zero vertical acceleration so that the flow is hydrostatic and parallel to the mean bed. 
As with most models for hydrostatic flows, we assume a small bed slope and refer to the flow parallel to the bed as being `horizontal' --- non-negligible bed slopes and their influence on morphological evolution are treated in detail in \S \ref{sec:bedslope}. 
A single sediment size and uniform bed porosity are considered. 
In each layer, the horizontal velocity, $u(x,t)$, and sediment concentration, $c(x,t)$, are assumed uniform with depth (see Figure \ref{fig_vertical_structure}), but varying in the streamwise direction and time. 
The bedload layer is assumed to have constant, arbitrary, vanishing thickness and variable density. 
This permits simulation of the (often ambiguously defined) bedload layer as a near-bed transport zone, whose sediment concentration can vary from zero for no sediment transport, to a maximum or saturation value. 
While typical 2-Layer models \cite[e.g.][]{Fraccarollo2002, Spinewine2005, Zech2008} aim to track the evolution of an interface dividing two layers with different but homogeneous densities (usually those of water and water-sediment mixture), in the present model such an interface is instead thought of as an imaginary line whose sole role is to delimit the near-bed region; in other words, it sets a tracking volume near the bed. 
The idea behind this assumption is to translate into the mathematical model the inherent ambiguity in the distinction between different modes of transport by fixing an arbitrary, near-bed layer concerned with bedload, and to compensate for this arguably restricting condition by relaxing the assumption of an upper layer composed of clear water (i.e. allow sediment to enter the upper layer). 
This key feature distinguishes the present model from other 2-Layer models, which is why we refer to it as a Quasi-2-Layer (Q2L) model. 

The model may simulate three different modes of transport:

\begin{itemize}
\item Mode 0: no sediment transport. For flow conditions below the threshold of sediment motion, both layers consist of pure water (Figure \ref{fig_vertical_structure}.b).

\item Mode 1: bedload only. Flow conditions are such that only the lower layer carries sediment and the upper layer consists of pure water (Figure \ref{fig_vertical_structure}.c). The model is primarily concerned with this type of transport.

\item Mode 2: total load. We relax the assumption of an upper layer consisting of clear water, such that at higher flow conditions, the bedload layer has reached a saturation point and sediment entrains into the upper layer, where it is treated as suspended load (Figure \ref{fig_vertical_structure}.d). Only low concentrations of suspended sediment are considered.\\
\end{itemize}

It is assumed that the fluid is always in motion, encompassing hydrodynamic conditions leading to the modes described above. 
Cases not modelled include: still fluid; sediment being present in the upper layer when the lower layer has not reached saturation; sudden entrainment caused by intense turbulence or a lateral source of sediment; and fine cohesive sediments (which may enter suspension mode without necessarily going through the bedload stage). 
Sheet-flow transport, where a distinct interface occurs between the lower transport layer and the upper pure-water layer, may be simulated as Mode 1. 
Flows carrying highly concentrated suspended loads are outside the scope of the present model, noting that different approaches \citep[e.g.][]{Fang2000, Rosatti2006,Chiodi2014} may be required in order to simulate such flows more accurately. 
Bedload-dominated problems are the main objective of the present model. 
However, the option of modelling low-concentration suspended load has been incorporated as a `safety valve' that allows the user to study a problem without \emph{a priori} certainty that only bedload transport will take place. 
To model entrainment and deposition, the bed erosion rate is estimated from the conservation of horizontal momentum at the bed interface, following \cite{Fraccarollo2002}.

\subsection{Governing equations} \label{sec:gov-eqs}

\subsubsection{Derivation} \label{sec:derivation}

\begin{figure}
\centering
\includegraphics[width=0.65\linewidth]{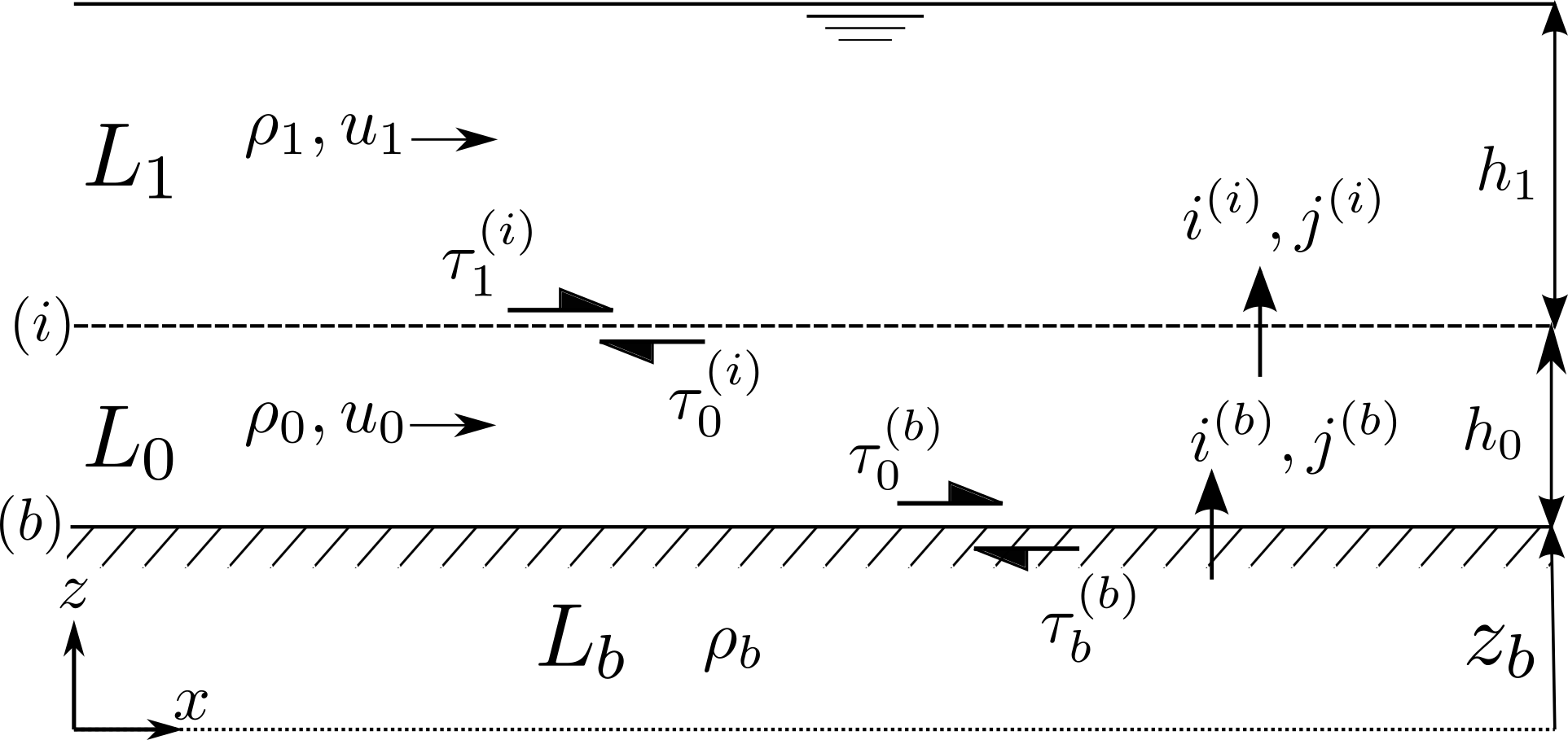}
\caption{Definition sketch for the Quasi-2-Layer model.} \label{fig_defQ2L}
\end{figure}

The governing equations are derived by: 
1) applying conservation laws to overall water-sediment mass in the upper layer, sediment mass in the upper layer, sediment mass in the lower layer, overall water-sediment horizontal momentum in the upper layer, overall water-sediment horizontal momentum in the lower layer, and sediment mass in the bed; 
2) considering mass and momentum exchanges between all three layers (bed included); and 
3) assuming thickness of the lower layer to be constant in space and time. 
Contribution of (a small) bed slope to momentum fluxes is considered. 
The following set of equations is obtained:
\begin{small}
	\begin{subequations} \label{eq_Q2Lfull}
		\begin{align}
		& \frac{\partial (\rho_1 h_1)}{\partial t} + \frac{\partial (\rho_1 h_1 u_1)}{\partial x} = i^{(i)}    \label{eq_Q2L-mass-up} \\ 
		& \frac{\partial (\rho_s  c_1 h_1)}{\partial t} + \frac{\partial (\rho_s  c_1 h_1 u_1)}{\partial x} = i_s^{(i)}   \label{eq_Q2L-sed-up} \\
		& \frac{\partial (\rho_s  c_0 h_0)}{\partial t} + \frac{\partial (\rho_s  c_0 h_0 u_0)}{\partial x} = i_s^{(b)} - i_s^{(i)}   \label{eq_Q2L-sed-low} \\
		& \frac{\partial (\rho_1 h_1 u_1)}{\partial t} + \frac{\partial }{\partial x} \left ( \rho_1 h_1 u_{1}^{2} + \frac{1}{2} \rho_1 g h_{1}^{2} \right ) + \rho_1 g h_1 \frac{\partial z_b}{\partial x} = j^{(i)}    \label{eq_Q2L-mom-up} \\
		& \frac{\partial (\rho_0 h_0 u_0)}{\partial t} + \frac{\partial }{\partial x} \left ( \rho_0 h_0 u_{0}^{2} + \frac{1}{2} \rho_0 g h_{0}^{2} \right ) + g h_0 \left [ \frac{\partial (\rho_1 h_1)}{\partial x} + \rho_0 \frac{\partial z_b}{\partial x} \right ] = j^{(b)} - j^{(i)}    \label{eq_Q2L-mom-low} \\
		& \frac{\partial  z_b}{\partial t} = -e^{(b)}   \label{eq_Q2L-vol-bed}
		\end{align}
	\end{subequations}
\end{small}
where subscripts `0' and `1' refer to the lower and upper layers, $L_0$ and $L_1$, respectively; $\rho$, $h$, and $u$ are the layer density, depth, and horizontal velocity, respectively; $g$ is gravitational acceleration; $z_b$ is bed level with respect to a fixed horizontal datum (subscript $b$ refers to the bed layer, $L_b$, assumed to be static; i.e. $u_b = 0$); $i$ and $j$ are net mass and momentum exchanges between layers (taken as positive in the upward direction) through interfaces denoted by superscripts $(i)$ and $(b)$ (see Figure \ref{fig_defQ2L}); $e^{(b)}$ denotes net water-sediment volumetric exchange between $L_b$ and $L_0$ (constant value of the bed bulk density, $\rho_b$, has been assumed). 
Note that the assumption of uniform velocity profiles implies the Boussinesq profile coefficient (commonly found in depth-averaged models) is equal to unity in the momentum balance equations, and so is not included here. 
A similar remark applies to the concentration profiles; i.e. a uniform, fully mixed profile, in conjunction with uniform velocity, yields a profile factor equal to unity in the mass conservation equations. 
The key assumption of $\partial h_0 / \partial t = \partial h_0 / \partial x = 0$ is implicit in \eqref{eq_Q2Lfull}. 
Observe that \eqref{eq_Q2L-sed-up} and \eqref{eq_Q2L-sed-low} imply that sediment particles transported by the fluid at the same stream-wise speed as the whole water-sediment mixture. 
In other words, $u_{sk} = u_k$ is assumed, where $u_{sk}$ represents the stream-wise velocity of sediment particles in layer $L_k$ ($k=0,1$). 
This is a sensible assumption for the upper layer, given that, for suspended load, sediment is expected to be transported at about the same speed of the flow \cite{Soulsby1997}. 
As for bedload, in Appendix \ref{appA}, a Lagrangian model for particle saltation is used to test the sensibility of this hypothesis, revealing that sediment transport rates predicted by the model remain virtually unaffected. 
Therefore, equivalence is pragmatically assumed between the sediment particle stream-wise velocity, $u_{s0}$, and the velocity of the corresponding water-sediment-mixture, $u_0$. 
(Such equivalence is also backed by experimental evidence; see \cite{Ni2015}.) 
In this first stage of model development, we try to retain simplicity whenever possible, acknowledging that additional modifications may be needed eventually to render the model more applicable to real, complex scenarios (see \S\S \ref{sec:model_validation}\ref{sec:experiment} and \ref{sec:conclusions}).

\subsubsection{Erosion rate and shear stresses} \label{sec:erosion_rate}

Following \cite{Fraccarollo2002}, the bed erosion rate is estimated from conservation of longitudinal momentum at the bed interface. 
Across interface $(b)$, momentum flux has to be single-valued, and so $j^{(b)}$ ought to be computed from variables at either side of the interface, yielding: $j^{(b)} = i^{(b)}  u_0 - \tau^{(b)}_{0} = - \tau^{(b)}_{b}$ (recall that $u_b = 0$), where $\tau^{(b)}_{0}$ is the shear stress exerted by the fluid on the bed surface and $\tau^{(b)}_{b}$ is the bed resistance, as depicted in Fig. \ref{fig_defQ2L} . 
The erosion rate, $e^{(b)} = i^{(b)} / \rho_b$, can then be estimated as:
\begin{equation} \label{eq_erosion}
e^{(b)} = \frac{1}{\rho_b \left | u_0 \right |} (\tau_{0}^{(b)} - \tau_{b}^{(b)}) .
\end{equation}
 
As stated previously, the present idealisation assumes the fluid is always in motion; however, should $u_0 = 0$ occur at some point in the domain at a certain time, the condition $e^{(b)} = 0$ for $u_0 = 0$ is imposed in order to avoid a mathematical error being introduced.

Herein, $\tau^{(b)}_{0}$ is estimated using a Ch\'{e}zy-type expression dependent on the squared velocity jump at the bed interface, $u_0$, and the bedload layer average density, $\rho_0$; namely:
\begin{equation} \label{eq_tau_0b}
\tau_{0}^{(b)} = C^{(b)} \rho_0 \left | u_0 \right | u_0 , 
\end{equation}
where $C^{(b)}$ is a friction coefficient, one of the main calibration parameters within the present model. 
By making $\tau^{(b)}_{0}$ dependent on $\rho_0$ (as opposed to the water density, $\rho_w$), the idea is to incorporate, even if crudely, the influence of the sediment-transport contribution to the total bed shear stress. 

Following \cite{Spinewine2005,Zech2008}, the bed interface is treated as a failure plane, such that the shear stress $\tau^{(b)}_{b}$ can be related to Terzaghi's effective normal stress through Coulomb's law, yielding:
\begin{equation} \label{eq_tau_bb}
\tau^{(b)}_{b} = \left \{  \tau_c +  [ h_1 (\rho_1 - \rho_w) + h_0 (\rho_0 - \rho_w)] g \tan \varphi  \right \} \left | u_0 \right | / u_0,
\end{equation}
where $\rho_w$ is the density of water, $\tau_c$ is the critical yield stress, obtained from the Shields curve, and $\varphi$ is the soil friction angle, taken as equal to the angle of repose. 
Note that both $\tau_c$ and $\varphi$ depend on sediment characteristics, such as the type of sediment, its density, and its particle diameter. 
The term $\left | u_0 \right | / u_0$ is included to ensure that $\tau^{(b)}_{b}$ acts as a resistive stress oriented opposite to the flow direction (Figure \ref{fig_defQ2L}). 
Comprehensive descriptions of the methodology and assumptions underpinning the derivation of shear stresses and erosion rate functions stated above are given by \cite{Spinewine2005, Fraccarollo2002, Abbott1979}.  

Flux of horizontal momentum at $(i)$ is also required to be single-valued regardless of whether variables from the top or bottom side of the interface are invoked. 
However, such a flux does not evolve freely, but instead depends on the bed erosion rate, $e^{(b)}$, thus requiring only one of the two shear stresses at interface $(i)$ (i.e. $\tau_1^{(i)}$ or $\tau_0^{(i)}$) to be computed --- the second becoming redundant. 
Arbitrarily, we choose to compute the shear stress at the bottom side of interface $(i)$, $\tau_0^{(i)}$. 
A Ch\'{e}zy-type expression provides a simple way to estimate $\tau_0^{(i)}$, consistent with the assumed vertical structure of the flow; namely:
\begin{equation} \label{eq_tau_i}
\tau_0^{(i)} = C^{(i)} \rho_1 \left | u_1 - u_0 \right | (u_1 - u_0) ,
\end{equation}
where $C^{(i)}$ is a second calibration coefficient, the first being $C^{(b)}$ in eq. \eqref{eq_tau_0b}. 

Closure relationships for shear stresses herein adopted prioritise simplicity and ought to be considered as exploratory measures, which should be revised in the future by comparison against high-quality experimental data.

\subsection{Inter-layer mass and momentum fluxes} \label{sec:interlayer-fluxes}

Starting from Mode 0, when bed material is initially eroded, $L_b$ and $L_0$ exchange \emph{water-sediment} mass. 
Conservation of volume within $L_0$ (i.e. $h_0 =$ constant) requires a compensating flux of \emph{water} between $L_0$ and $L_1$. 
This is true until $L_0$ gets saturated with sediment (i.e. $c_0$ has attained its maximum permitted value, $c_{0 \, \textup{mx}}$), in which case the \emph{water-sediment} mass mixture eroded/deposited from/onto $L_b$ is compensated by an equal amount of \emph{water-sediment} mass exchanged between $L_0$ and $L_1$. 
Therefore, the net mass fluxes through interfaces $(b)$ and $(i)$ are expressed as:

\begin{equation} \label{eq_ib}
i^{(b)} = \begin{cases}
0 & \text{if} \; \; \; e^{(b)} <0 \; \; \; \textup{and} \; \; \; c_0=c_1=0 \\ 
\rho_b e^{(b)} & \text{else}
\end{cases}
\end{equation}

and

\begin{equation} \label{eq_ii}
i^{(i)} = \begin{cases}
0 & \text{if} \; \; \; i^{(b)}=0  \\ \\
\rho_w e^{(b)} & \text{if} \; \; \; c_0 \ne 0  \; \; \; \text{and} \; \; \; c_1=0 \\ \\
\rho_b e^{(b)} & \text{if} \; \; \; c_0 = c_{0 \, \textup{mx}}  \; \; \; \text{and} \; \; \; c_1 \ge 0
\end{cases} \; \; ,
\end{equation}

The above expressions underpin the logical requirement that no sediment may be deposited onto the bed when operating as Mode 0 (where $\tau_0^{(b)} < \tau_b^{(b)} \Rightarrow e^{(b)} < 0$, would predict deposition even if no sediment is present in $L_0$).

The corresponding \emph{sediment} mass fluxes are the sediment components of the total water-sediment mass exchanges, and are expressed as:
\begin{equation} \label{eq_ib_s}
i^{(b)}_s = c_b \, \frac{\rho_s}{\rho_b} \, i^{(b)} ,
\end{equation}

and

\begin{equation} \label{eq_ii_s}
i^{(i)}_s = \left\{\begin{matrix}
0 & \textup{if} \; \; \; c_0 < c_{0 \, \textup{mx}} \; \; \; \textup{and} \; \; \; c_1=0 \\ \\
c_b \, \rho_s \, e^{(b)} & \textup{if} \; \; \; c_0 = c_{0 \, \textup{mx}}  \; \; \; \textup{and} \; \; \; c_1 \ge 0
\end{matrix}\right. \; \; .
\end{equation}
where $\rho_s$ is the density of sediment.

Exchange of horizontal momentum between layers takes place when mass crosses the interface from one layer to an adjacent layer. 
This occurs at both interfaces $(i)$ and $(b)$ when $i^{(b)} \ne 0$ (and hence $i^{(i)} \ne 0$). 
As mentioned before, momentum fluxes at such interfaces are to be single-valued. 
However, at the bed interface, we compute the horizontal momentum flux from variables at the upper part of interface $(b)$; namely:
\begin{equation} \label{eq_jb_final} 
j^{(b)} = i^{(b)} u_0 - \tau_{0}^{(b)}.
\end{equation}

This is because \eqref{eq_jb_final} implies the physically meaningful condition that $L_0$ faces solely a resistive bed friction proportional to the square of its velocity when the model is operating as Mode 0. 
Moreover, analysis of eq. \eqref{eq_jb_final} enables correct interpretation of the momentum flux, $j^{(b)}$. 
This is important because, as remarked upon by \cite{Iverson2015}, such a flux has been occasionally reported in the literature to yield the seemingly illogical conclusion that $L_0$ gains momentum due to mass crossing the interface $(b)$ from a state originally at rest (and hence with no initial momentum). 
Under no sediment transport conditions, the total resistance encountered by the flow is $\left | \tau_{0}^{(b)} \right |  = \left | \tau_{b}^{(b)} \right |$. 
However, if bed material is eroded, the frictional momentum acting at the bottom of $L_0$ is reduced by a factor of $- \left | i^{(b)} u_0 \right | $, and thus any apparent gain of momentum of $L_0$ is in actuality a reduction of the diffusive momentum or basal friction \cite{Iverson2015}.

As with the bed, net flux of momentum at $(i)$ can be evaluated from variables at either side of the interface. 
However, unlike $(b)$, for $(i)$ there is no preferential candidate based on physical significance, and so to ensure consistency with the previous arbitrary parameterisation of $\tau_0^{(i)}$ (eq. \ref{eq_tau_i}), variables from $L_0$ are invoked, yielding:
\begin{equation} \label{eq_ji}
j^{(i)} = i^{(i)} u_0 - \tau_0^{(i)} .
\end{equation}

Eqs. \eqref{eq_tau_0b}-\eqref{eq_ji} close the set of governing equations given by \eqref{eq_Q2Lfull}.

\section{Model validation} \label{sec:model_validation}

We devote this section to analysis of the model and comparison against established theory and experiments on sediment transport rates and morphodynamics. 
Especial emphasis is given to bedload-dominated problems, since these are the main aim of the present model, as previously remarked upon. 
Further details of the mathematical treatments, proofs and derivations that follow can be found in \cite{MaldonadoT2015}.

\subsection{Analytical solution for bedload} \label{sec:bedload-analyt}

We consider steady uniform flow over an erodible bed with bedload transport exclusively present ($c_0 \le c_{0 \, \textup{mx}}$ and $c_1 = 0$). 
This case permits derivation of an analytical solution to the Q2L model, which can then be used to compare the present model against bedload theory, including validation against empirical formulae. 
The volumetric bedload transport rate, $q_b$, is evaluated as: $q_b = h_0 c_0 u_0$. 
 Note that the \emph{sediment} bedload rate should strictly be computed as $q_b = h_0 c_0 u_{s0}$. However, this would require an additional equation relating $u_{s0}$ to the model output $u_0$. 
Appendix \ref{appA} proves that the assumption $u_0 \approx u_{s0}$ appears sensible from quantitative and pragmatic perspectives and is thus adopted herein. 

For steady uniform flow that is initially above the threshold of sediment motion, equilibrium conditions for sediment transport are expected to be reached eventually. 
Given that $h_0$ is a constant within the present model, equilibrium-state values have to be found solely for $c_0$ and $u_0$ in order to compute $q_b$. 
Once bed erosion has initiated, steady sediment transport conditions can only occur once $e^{(b)}$ decreases to zero. This happens when both $c_0$ and $u_0$ have reached certain values, $c_{0 \, \text{eq}}$ and $u_{0 \, \text{eq}}$, respectively, such that the force exerted by the water-sediment flow on the bed surface equals its resistance to erosion. 
In other words, $\tau_{0}^{(b)} = \tau_{b}^{(b)}$ has to be verified in order for $e^{(b)} = 0$ to occur. 
Hence, $\tau_0^{(b)} = \tau_b^{(b)} \Rightarrow \tau_0^{(b)}  = \tau_c + (\rho_{0} - \rho_w) h_0 g \tan \varphi$; from which an expression for the equilibrium bedload layer density, $\rho_{0 \, \text{eq}}$, and thus $c_{0 \, \text{eq}}$, can be found. 
Then, an expression for $ u_{0 \, \text{eq}}$ simply follows from \eqref{eq_tau_0b}, allowing calculation of the bedload transport rate, equal to $h_0 \, c_{0 \, \text{eq}} \, u_{0 \, \text{eq}}$, from:
\begin{equation} \label{eq_qb-mode1-ftau}
q_b = \frac{(\tau_0^{(b)} - \tau_c )}{ (\rho_s - \rho_w) g \tan \varphi} \left( \frac{\tau_0^{(b)}}{\rho_{0 \, \text{eq}} \; C^{(b)}} \right)^{1/2} .
\end{equation}

Note that bedload is independent of the second calibration parameter, $C^{(i)}$. 

\subsection{Mathematical agreement with bedload formulae} \label{sec:mathe_agree}

Inspection of \eqref{eq_qb-mode1-ftau} reveals that the bedload transport rate predicted by the present model follows the general form:
\begin{equation} \label{eq_qb_analyt}
q_b = A' \left( \tau_0^{(b)} - \tau_c \right) \left( \tau_0^{(b)} \right)^{1/2} ,
\end{equation}
where $A' = \left[ (\rho_s - \rho_w) g \tan \varphi \right]^{-1} \left( \rho_{0 \, \text{eq}} C^{(b)} \right)^{-1/2} $. This is in agreement with several empirical and semi-empirical bedload formulations \cite[e.g.][]{Bagnold1963, Nielsen1992, Soulsby1997, Yalin1963}, which can be written in the generic form: $q_b = F (\tau - \tau_c) \, \tau^{1/2}$, where $F$ is an expression often taken as a constant obtained from a best-fit curve to laboratory data, and $\tau$ ($\cong \tau_0^{(b)}$) represents the bed shear stress. 
In Appendix \ref{appA}, it is shown that by computing bedload as $q_b = h_0 c_0 u_{s0}$, and relating $u_{s0}$ to $u_0$ by means of a Lagrangian saltating particle model, $q_b$ instead follows the form $q_b = A' (\tau - \tau_c) (\widehat{d} \, \tau^{1/2} - \widehat{c} \, \tau_c^{1/2} )$ (where $\widehat{c}$ and $\widehat{d}$ are calibration coefficients), which is also in agreement with various empirically derived expressions \cite[e.g.][]{Madsen1991, Ashida1972}.

Also note that for sheet flow conditions (a special regime of bedload), where $\tau_0^{(b)} \gg \tau_c \Rightarrow \tau_0^{(b)} - \tau_c \approx \tau_0^{(b)}$, it can readily be shown from \eqref{eq_qb_analyt} and \eqref{eq_tau_0b} that the model predicts $q_b \propto u_0^3$, in agreement with established sheet flow theory.

\subsection{Validation (empirical bedload formulae)} \label{sec:bedload-validation}

\begin{figure} 
	\begin{center}
		\begin{minipage}{0.95\linewidth}
			\subfigure[medium sand]{
				\resizebox*{0.45\linewidth}{!}{\includegraphics{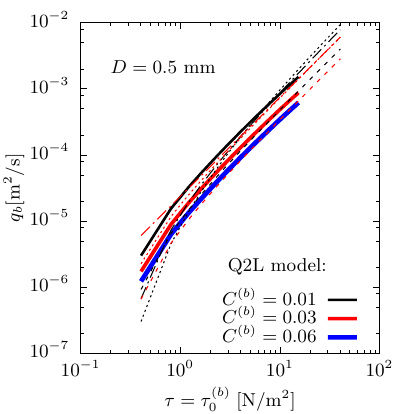}} \label{fig_validation_dim_a} } \hspace{5pt} 
			\subfigure[coarse sand]{
				\resizebox*{0.45\linewidth}{!}{\includegraphics{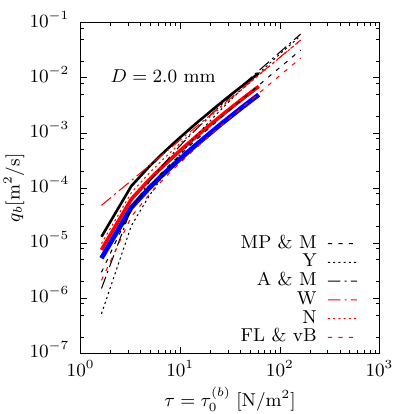}} \label{fig_validation_dim_b} }
			\caption{Comparison between bedload rates as functions of bed shear stress predicted by the Q2L model (continuous lines) against six empirical and semi-empirical expressions (broken lines), for two particle diameters. [Acronyms defined in corresponding paragraph; see \ref{sec:bedload-validation}]}
			\label{fig_validation_dim}
		\end{minipage}
	\end{center}
\end{figure}

Fig. \ref{fig_validation_dim} compares model predictions (eq. \ref{eq_qb_analyt}) against corresponding values from popular empirical bedload formulae by \cite{MPM1948} (MP \& M), \cite{Yalin1963} (Y), \cite{Ashida1972} (A \& M), \cite{Wilson1966} (W), \cite{Nielsen1992} (N) and \cite{FernandezLuque1976} (FL \& vB). 
Two particle diameters, $D$, are considered: 0.5 and 2.0 mm. These particle sizes are chosen because bedload is likely to be the main mode of transport for $D>0.3$ mm \cite{Soulsby2005}.  Parameter values are $s \equiv \rho_s / \rho_w =2.65$,  $\varphi = 32.1^{\circ}$, and $h_0 = 10D$ (more on $h_0$ in \ref{sec:onh0}). 
Three values of $C^{(b)}$ are investigated; namely, 0.01, 0.03 and 0.06.

The present model predictions fall within the band of estimates delimited by the empirical formulae considered; this band is representative of the well-known uncertainty in the quantification of bedload. 
Here, the model predictions are truncated where $c_0 = c_{0 \, \text{mx}}$ is reached. 
Experiments by \cite{Spinewine2005} show that values of sediment concentration within the bedload layer tend to be confined to the range $[0.21, 0.25]$, and so a slightly larger value of $c_{0 \, \text{mx}} = 0.3$ is selected in order to extend the model prediction curves, noting that such a selection purely acts as the Mode 1 limit; in other words, the larger the value of $c_{0 \, \text{mx}}$, the longer the model can operate as bedload-only. 
Comparison between Figures \ref{fig_validation_dim_a} and \ref{fig_validation_dim_b} demonstrates that the overall behaviour of the Q2L model in relation to established empirical formulae is independent of particle size, within the range of parameters considered. 
The recommended value of $C^{(b)}$ depends on the reference formula, but the overall behaviour of the model matches empirical predictions, confirming the mathematical agreement discussed above. 

\subsection{Bedload layer thickness, $h_0$} \label{sec:onh0}

\begin{figure}
\centering
\includegraphics[width=0.6\linewidth]{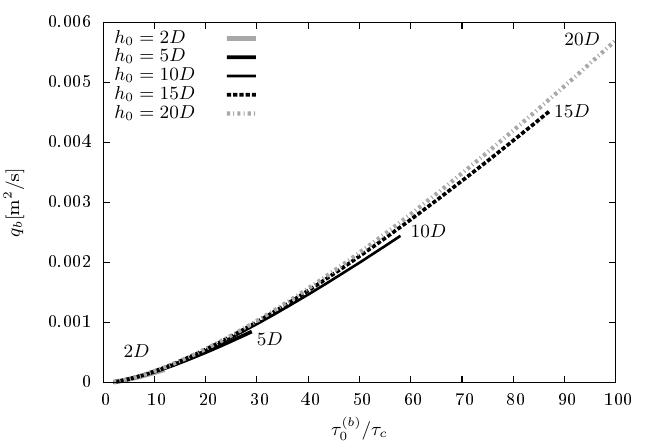}
\caption{Bedload predicted by the Q2L model as a function of the ratio of bed shear stress to critical shear stress (or transport stage) for different values of $h_0$.} 
\label{fig_h0}
\end{figure}

Except for certain specific cases, such as sheet flow, a reliable, general method does not exist by which to determine the thickness of the bedload layer. 
Here, we prescribe an arbitrary, yet realistic, thickness of the bedload layer (i.e. $h_0$) that represents the \emph{vicinity} of the bed where bedload occurs.  
Fig. \ref{fig_h0} investigates the sensitivity of the bedload predicted by the model to values of $h_0$ in the range of $[2D \, , \, 20 D]$. 
This range is selected noting estimates of bedload layer thickness by \cite{Einstein1950} and \cite{vRijn1984}. 
The particle diameter is 1.0 mm. 
The influence of the arbitrary $h_0$ on the predicted $q_b$, for a given $\tau_0^{(b)} / \tau_c$, is small. 
For $\tau_0^{(b)} / \tau_c = 20$ (beyond which sheet flow is expected), discrepancies between the curves are negligible. 
The most significant effect of the selected value of $h_0$ is the variation in range of model validity when operating as Mode 1. 
The curves are plotted up to the point where $c_0 = c_{0 \, \text{mx}} = 0.3$, beyond which sediment is considered (within the framework of the present model) to be transported as suspended load (Mode 2). 
A larger value of $h_0$ allows the model to operate as Mode 1 over a wider range of $\tau_0^{(b)}$. 
Note that a more conservative (lower) value of $c_{0 \, \text{mx}} $, such as $0.21 \lesssim c_{0 \, \text{mx}} \lesssim 0.25$ following \cite{Spinewine2005}, would further minimise the discrepancies between the curves in Fig. \ref{fig_h0} by shifting their truncation point (end of Mode 1) to the left.

From Fig. \ref{fig_h0}, it can be observed that $h_0 \approx 10 D$ yields a relatively wide range of validity for Mode 1, up to  $\tau_0^{(b)} \approx 60 \tau_c$. 
By use of a two-phase model, \cite{Chiodi2014} arrive at the conclusion that the thickness of the bedload layer is $\sim 10 D$. 
Therefore, the value $h_0 = 10 D$ is selected as the default unless otherwise stated.

\subsection{No transport and total load (Modes 0 and 2)} \label{sec:mode2}

Although the Q2L model is primarily designed for bedload-dominated problems, analysis of the other two possible modes of transport (i.e. no transport and total load) can lead to further useful insights. 
For example, the simple case of steady uniform flow below the threshold of motion can be invoked to prove that, in general, $C^{(b)}$ is not equal to (in fact, is larger than) typical bed friction coefficients derived from Ch\'{e}zy or Manning formulae, which are employed in standard 1-Layer hydrodynamic models. 
Moreover, under these conditions, values for the ratio $C^{(b)} / C^{(i)}$ can be proposed using either an assumed or a measured vertical velocity profile. 
For instance, if the flow velocity profile follows a power law as suggested by \cite{Soulsby1997}, an expression for $C^{(b)} / C^{(i)}$ as function of $h_1$ and $h_0$ can be obtained, which may be useful (see below).

\begin{figure}
\centering
\includegraphics[width=0.6\linewidth]{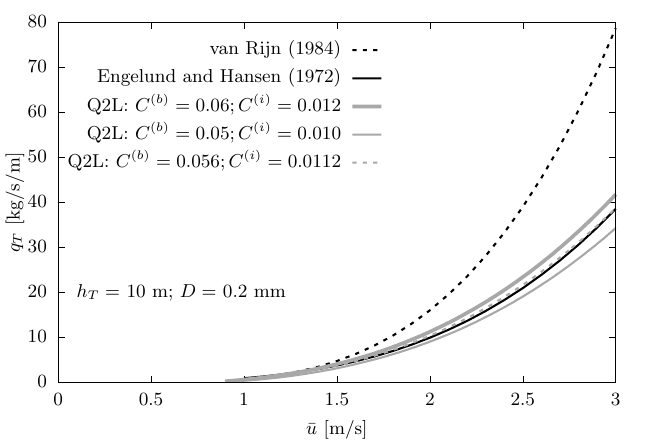}
\caption{Comparison between the Q2L model predictions and empirical estimates of total sediment transport , $q_T$, as function of flow depth-averaged velocity, $\bar{u}$. Three values of $C^{(b)}$ are considered; the ratio of $C^{(b)} / C^{(i)} = 5$ is fixed.} 
\label{fig_total_transport_1}
\end{figure}

Of more interest is the analysis of Mode 2 (total load), which arises from relaxing the assumption of an upper layer consisting always of pure water, as a form of compensation for the arguably limiting constraint of a fixed bedload layer thickness. 
The user does not need to be concerned about violating the bedload-only condition during the simulation, but sediment may enter the upper layer at times and locations where flow is sufficiently fast, so long as near-bed transport dominates. 
In this way, Mode 2 operates as a `safety valve' that adds flexibility to the model. 
To demonstrate the potential of the model to deal with total load, an analytical solution for Mode 2 has been derived (details not presented here, for brevity), and its results compared against empirical expressions for total transport proposed by \cite{Engelund1967} and \cite{vRijn1984b}. 
Let $h_T \equiv h_0 + h_1 =$ 10 m, $D = $ 0.2 mm (uniform sediment), and $h_0 = 10D$. 
Unlike the bedload-only case, the analytical solution for total load requires both tuning parameters $C^{(b)}$ and $C^{(i)}$. 
The additional degree of freedom for calibration (with respect to Mode 1) can be removed by proposing a value for the ratio $C^{(b)} / C^{(i)}$ based on hydrodynamic considerations, as mentioned at the beginning of this section. 
For values of $h_1$ and $h_0$ considered, $C^{(b)} / C^{(i)} \approx 5$ can be prescribed. 
Fig. \ref{fig_total_transport_1} compares predictions by the selected empirical formulae and the present model for $C^{(b)} = $ 0.05, 0.056, 0.06 and constant $C^{(b)} / C^{(i)} = 5$. 
The Q2L model fits the formula by \cite{Engelund1967} better than that of \cite{vRijn1984b}. 
For $C^{(b)} = 0.056$, the agreement achieved between the model prediction and that by \cite{Engelund1967} is outstanding over the range of parameters studied. 
The model predictions shown in Fig. \ref{fig_total_transport_1} are for a bedload layer saturation value of $c_{0 \textup{mx}} = 0.25$, following \cite{Spinewine2005}. 
Further studies (not included here) demonstrate that the model shows little sensitivity to the selection of $c_{0 \textup{mx}}$, over the realistic range $0.21 \lesssim c_{0 \, \text{mx}} \lesssim 0.25$, with discrepancies between curves vanishing for large flow velocities. 
Maximum values of $c_1$ for curves shown in Fig. \ref{fig_total_transport_1} are $\sim 0.001$, hence, $c_0 \gg c_1$, thus verifying the condition that bedload is the predominant mode of transport. 

It is not intended that Fig. \ref{fig_total_transport_1} be used to promote use of the model for problems dominated by suspended transport. 
Instead, Fig. \ref{fig_total_transport_1} merely indicates the potential of the model to cope with scenarios where complex transport patterns occur, where the distinction between bedload and suspended load is unclear. 
Such potential is worth further exploration, but it is outside the scope of the present paper, which is on bedload-governed cases. 

\subsection{Validation (morphodynamic experiment)} \label{sec:experiment}

\begin{figure} 
	\centering
	\includegraphics[height=5.5cm]{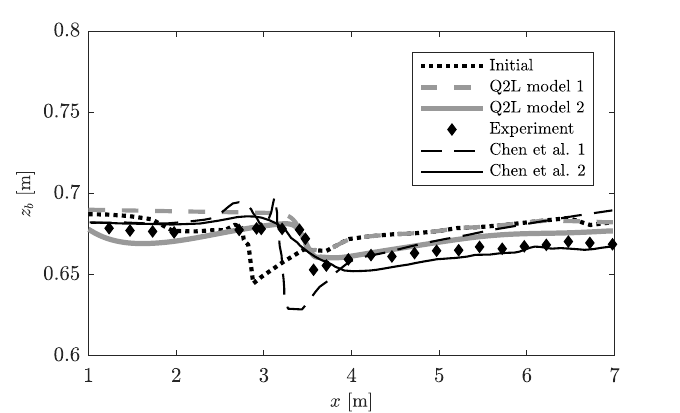}
	\caption{Evolution of a mining pit profile after 3 hours. Comparison between measurements by \cite{Lee1993} and predictions by the present model and that of \cite{Chen2010} (Chen et al.). [Legend is explained in corresponding paragraph; see \ref{sec:experiment}]} 
	\label{fig_q2l_pit}
\end{figure}

To test the model we select for comparison an experiment carried out by \cite{Lee1993}. 
This experiment studies the migration of a mining pit due to bedload driven by a steady current. 
For reasons that become evident in \S \ref{sec:bedslope}, we focus on the second stage of the experiment, referred to by \cite{Lee1993} as the `diffusion period'. 
Fig. \ref{fig_q2l_pit} illustrates this comparison. 
For reference, predictions by \cite{Chen2010} (Chen et al.) are also included. 
In \cite{Chen2010}, a CM model is employed (i.e. hydrodynamic and morphological models coupled via a sediment transport empirical formula) with a bed-update equation modified by inclusion of an adaptation length (the distance it takes bedload to adjust from a non-equilibrium state to an equilibrium one). 
Fig. \ref{fig_q2l_pit} includes predictions by \cite{Chen2010} for two values of this adaptation length; namely, 1 cm (labelled `Chen et al. 1') and 2 cm (`Chen et al. 2' --- their reported best fit). 
Two predictions by the present model are also shown. 
In the first case (`Q2L model 1'), values of the calibration parameters $C^{(b)} = 6.55 \times 10^{-3}$ and $C^{(i)} = 4.50 \times 10^{-2}$ are used, which yield the correct migration speed of the pit. 
However, the value reported by \cite{Lee1993} of upstream transport stage (ratio of bed shear stress to critical shear stress) is not replicated. 
In fact, when we aim to reproduce such a value ($\approx 1.77$), the predicted migration of the pit is significantly faster than the one reported. 
This has motivated the introduction of a further calibration parameter, similar to that of \cite{Chen2010}, in the bed-evolution equation \eqref{eq_Q2L-vol-bed}, leading to: $\partial z_b / \partial t = - \eta_e e^{(b)}$. 
(See discussion in \S \ref{sec:bedslope}\ref{sec:dif_exp_CM} regarding the abnormally low value of $F_*$ required to reproduce the reported experimental settings in the context of a CM model.) 
Here, $\eta_e$ deals with the irregularity in shape and size of the bed material present in the experiment, not accounted for in the model derivation (that assumes perfectly uniform sediment), which impact the bed's packing fraction and thus the vertical distribution of its bulk density, $\rho_b$. 
(See Appendix \ref{AppB} for further details.) 
No detailed information on the bed composition is given in \cite{Chen2010} that could aid the development of a sophisticated representation of $\eta_e$, and so a constant value is assumed as a first step. 
A value of $\eta_e =  0.15$ leads to the curve labelled `Q2l model 2' in Fig. \ref{fig_q2l_pit}, where not only the correct pit migration speed is achieved, but also the reported value of the transport stage is replicated. 

The present model without any modification (Q2L model 1) is able to predict the correct migration speed and final depth of the pit. 
However, from a qualitative perspective, significant discrepancies occur, with the overall bed level higher than observed, precisely because the low transport stage upstream ensures that no erosion takes place downstream of the pit (flow is from left to right in the plot). 
The second prediction by the model (Q2L model 2) gives better results, especially from a qualitative viewpoint. 
Note that this curve properly predicts the observed inflection point in the bed profile upstream of the pit, and the steepness of the upstream face of the pit. 
The latter condition is not predicted by \cite{Chen2010}. 
Instead, the `Chen et al. 1' results develop an unrealistic peak upstream of the pit, whereas the `Chen et al. 2' results introduce diffusion that avoids development of the unrealistic peak, but at the cost of over-diffusing the pit profile. 

The present experimental comparison indicates the potential of the model to deal with real-world morphodynamic problems, achieving predictions of quality at least comparable to previous studies. 
It, however, also highlights the need for future investigation into the idealised assumptions underpinning the derivation of the model. 
In the next section, the model is utilised to derive an expression for morphological diffusivity that can be incorporated into CM models for application to cases similar to the one here studied.

\section{Bed slope influence on bedload: morphological diffusion} \label{sec:bedslope}

Under scenarios involving steep local bed slopes, such as mountain streams and certain beaches, the gradient of the bed elevation may play an important role in sediment transport processes and morphological evolution \cite{Bayazit1983,Amoudry2011} as confirmed through laboratory experiments \cite[e.g.][]{Smart1984, Damgaard1997, Dey2001}. 
Additionally, inclusion of the effect of bed slope on bedload may prevent the evolution of unrealistic morphological oscillations without the need to resort to specialised numerical techniques, while enhancing the physical significance of the model -- bed gradients help diffuse out spurious bed features \cite{Johnson2002}, thus the term `morphological diffusivity'. 
Methods used to account for the influence of bed slope include: addition of a slope-related diffusivity term to a sediment transport formula, which typically translates into an additional calibration parameter \cite{Johnson2002}; semi-empirical models based on Bagnold's ideas \cite[e.g.][]{Bagnold1963, Bailard1981a, Kovacs1994}; formulae explicitly derived for sloping beds, which often imply a significant degree of empiricism or complexity \cite[e.g.][]{Smart1984, Chiari2010, Parker2003}; and modification of the threshold of motion for sloping beds by inclusion of the weight of the particle at rest. 
In this section, we propose an analytically derived expression for morphological diffusivity that can be incorporated into a CM model for use in bedload-dominated problems.

\subsection{Quantification of the bed slope influence on bedload}

Here, we assess the bed slope influence on bedload using the ratio of bedload transport on a sloping bed to the bedload that would occur in a horizontal channel for the same bed shear stress. 
This is expressed as:
\begin{equation} \label{eq_slope-infl-def}
\Pi_\beta \equiv \frac{ q_b |_{\tau_c = \tau_{c \beta}, \tau = a_0} }{ q_b |_{\beta=0 \therefore \tau_c = \tau_{c h}, \tau = a_0} } ,
\end{equation}
where $a_0$ is a given value of $\tau$, $\beta$ is the bed slope angle, $\tau_{c \beta}$ represents the threshold of particle motion for a bed of arbitrary slope, and $\tau_{c h}$ is the threshold for a horizontal bed. 
Both quantities are related by incorporating the effect of gravity on particles at rest on a given slope, such that:
\begin{equation} \label{eq_ratio_tau_slope}
\frac{\tau_{c \beta}}{\tau_{c h}} = \frac{ \sin(\varphi + \beta) }{ \sin \varphi } .
\end{equation}  

Hence, using the analytical solution derived for bedload under steady uniform flow, \eqref{eq_qb-mode1-ftau}, the bed slope influence, $\Pi_\beta$, can be rewritten (replacing $\tau_0^{(b)}$ by $\tau$) as:
\begin{align}
\Pi_\beta &=  \left (  \frac{ \rho_{0 h} }{ \rho_{0 \beta} }  \right )^{1/2} \frac{ (\tau - \tau_{c \beta}) }{ (\tau - \tau_{c h}) } \nonumber \\
&= \Pi_1^{1/2} \, \Pi_2 , \label{eq_ratio_Pi}
\end{align}
where $\Pi_2 \equiv (\tau - \tau_{c \beta}) / (\tau - \tau_{c h})$ and $\Pi_1 \equiv \rho_{0 h} / \rho_{0 \beta}$. 
(Subscripts $\beta$ and $h$ denote sloping and horizontal beds, respectively.) 
Invoking the principle of equilibrium sediment transport conditions, such that $\tau = \tau_b^{(b)} \Rightarrow e^{(b)} = 0$, as described in \S \ref{sec:model_validation}, expressions for $\rho_{0 h}$ and $\rho_{0 \beta}$ can be obtained, yielding:
\begin{equation} \label{eq_ratio_qb_slp_ftau2}
\Pi_\beta = \left (  \frac{ \tau + \rho_w g h_0 \tan \varphi - \tau_{c h} }{ \tau + \rho_w g h_0 \tan \varphi - \tau_{c \beta} } \right )^{1/2} \left (   \frac{ \tau - \tau_{c \beta} }{ \tau - \tau_{c h} } \right ) .
\end{equation}

\begin{figure}
 	\centering
 	\includegraphics[height=5.5cm]{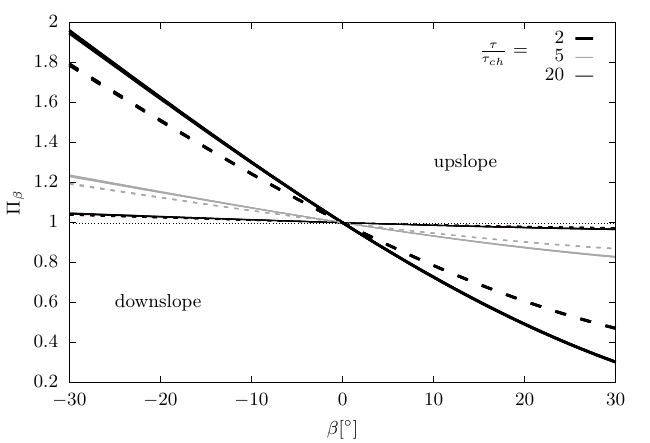}
 	\caption{Bed slope influence on bedload as a function of bed slope angle ($\beta > 0$ represents adverse slope). Twenty-four curves (generated from the combinations of different values of $\varphi$, $\tau / \tau_{c h}$, $h_0$ and $D$ considered) are plotted, grouped into six families corresponding to the six combinations of $\varphi$ and $\tau / \tau_{c h}$ analysed. Solid(dashed) lines correspond to $\varphi=$ 31(37)$^{\circ}$. } 
 	\label{fig_slope_influence_pi}
\end{figure}

The above equation depends on other variables besides the bed slope angle, $\beta$; namely: $\tau_{c h}$, $\tau$, $\varphi$, and $h_0$. 
Hence, a sensitivity analysis is undertaken using the following data (typical of a bedload-dominated scenario): $D = $ 0.5 and 2.0 mm; $s=2.63$; $h_0 = 5D$ and $15D$; $\tau / \tau_{c h}$ (not to be confused with $\tau / \tau_{c \beta}$) $=$ 2, 5 and 20; and $\varphi = $ 31 and 37$^{\circ}$. 
The results are plotted in Fig. \ref{fig_slope_influence_pi}, where it can be seen that the bed slope influence seems to be governed by the bed shear stress and the angle of repose, but insensitive to the selection of $D$ and $h_0 / D$ within the ranges of values considered. 
The bed slope influence also vanishes ($\Pi_\beta \rightarrow 1$) for large values of the bed shear stress, in agreement with experimental findings reported by \cite{Damgaard1997}. 
This is confirmed mathematically by taking the limit of eq. \eqref{eq_ratio_qb_slp_ftau2} as $\tau \to \infty$.
 
Further study on the sensitivity of $\Pi_\beta$ to $D$ and $h_0$ (not included here for brevity, but see \cite{MaldonadoT2015}), demonstrates that for all combinations of values considered, $\Pi_1^{1/2}$ falls in the range of $1 \pm \sim 1 \%$, which justifies the assumption of $\Pi_1^{1/2} \approx 1$, hence simplifying the bed slope influence as follows:
\begin{equation} \label{eq_ratio_qb_slp_ftau_approx}
\Pi_\beta \approx \frac{ \tau - \tau_{c \beta} }{ \tau - \tau_{c h} } . 
\end{equation}

The above equation is solely dependent on $\beta$, $\varphi$ (through eq. \ref{eq_ratio_tau_slope}) and the bed shear stress, $\tau$. 
It is also worth remarking that both the exact and approximate expressions for quantifying the bed slope influence on bedload are independent of the calibration parameter $C^{(b)}$.
 
 \subsection{The morphological diffusion}

A common way of accounting for the influence of bed slope is to modify the bedload formulae originally derived for horizontal channels by adding a term that promotes(inhibits) sediment transport in down(up)-sloping beds \cite{Johnson2002, Watanabe1988, Bailard1981a}. 
Such a term is proportional to the bed slope, and can be added as follows:
 \begin{equation} \label{eq_qb_add_diff}
 q_{b \beta} = q_{b h} + \varepsilon_\beta |q_{b h}|  S_b ,
 \end{equation}
where $S_b \equiv \partial z_b / \partial x \equiv \tan \beta$ is the bed slope, and $\varepsilon_\beta$ is a proportionality parameter related to morphological diffusion; $\varepsilon_\beta$ is often taken as an additional tuning parameter in morphodynamic models \cite{Johnson2002, Watanabe1988}. 
Assuming, for convenience, positive unidirectional flow (i.e. $\bar{u} > 0 \Rightarrow q_{b h} > 0$), eq. \eqref{eq_qb_add_diff} can be rewritten as:
\begin{equation} \label{eq_qb_linear}
 \frac{q_{b \beta}}{q_{b h}} = 1 + \varepsilon_\beta  S_b .
\end{equation}

The ratio $q_{b \beta} / q_{b h}$ (consistent with our definition of bed slope influence, $\Pi_\beta$; see eq. \ref{eq_slope-infl-def}) varies linearly with $S_b$; the parameter $\varepsilon_\beta$ is the slope of the line (with $y$-intercept equal to 1). 
From Fig. \ref{fig_slope_influence_pi}, it can be observed that the non-linear expression derived for $\Pi_\beta \equiv q_{b \beta} / q_{b h}$ exhibits quasi-linear behaviour consistent with \eqref{eq_qb_linear} over a relatively wide range of $\beta$. 
Thus, an expression for $\varepsilon_\beta$ can be proposed based on the bed-slope influence predicted by the present model. 
 
We extract analytically a value of $\varepsilon_\beta$ by obtaining the slope of the line tangent to the curve $\Pi_\beta$ vs $S_b$ (as given by the approximation \ref{eq_ratio_qb_slp_ftau_approx}) at the origin $S_b = \tan \beta = \beta = 0$; namely:
\begin{equation} \label{eq_morpho-diff}
\varepsilon_{\beta} = \frac{\partial \Pi_{\beta} }{\partial S_b} \Big |_{\beta = 0} . 
\end{equation}

By rewriting \eqref{eq_ratio_qb_slp_ftau_approx} as:
\begin{equation*}
\Pi_\beta \approx \frac{\tau - \tau_{c h} (\tau_{c \beta} / \tau_{c h}) }{\tau - \tau_{c h}} \, ,
\end{equation*}
it is evident that $ \tau_{c \beta} / \tau_{c h} $ depends on $S_b$($= \tan \beta$) through eq. \eqref{eq_ratio_tau_slope}. Hence,
\begin{equation*}
\frac{\partial (\tau_{c \beta} / \tau_{c h} )}{\partial S_b} = \frac{\partial }{\partial \tan \beta} \left [ \frac{\sin(\varphi + \beta)}{\sin \varphi} \right ] = \frac{\cos (\varphi + \beta)}{\sin \varphi} \cos^2 \beta . \label{eq_dtaus_dtan}
\end{equation*}

Invoking \eqref{eq_morpho-diff}, the morphological diffusivity is thus given by:
\begin{equation} \label{eq_diff_par_ftau}
\varepsilon_{\beta} = - \left ( \frac{ \tau_{c h}}{\tau - \tau_{c h}}  \right )\left ( \frac{1}{\tan \varphi} \right ) .
\end{equation}

\begin{figure}
	\centering
	\includegraphics[height=5.5cm]{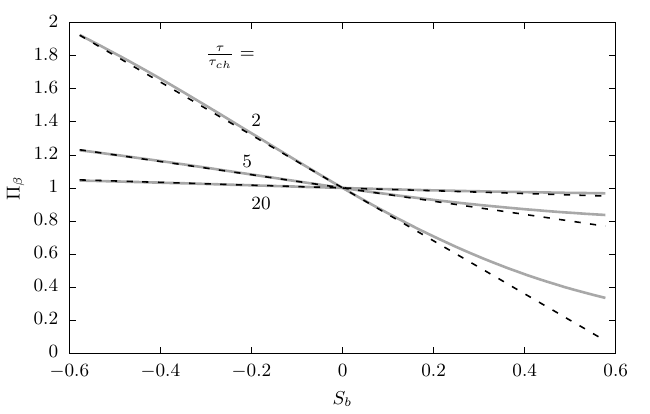}
	\caption{Bed slope influence as a function of bed slope for values of $\tau / \tau_{c h} =$ 2, 5 and 20; $\varphi = 32^{\circ}$. Comparison between exact solution given by eq. \eqref{eq_ratio_qb_slp_ftau2} (solid lines) and linear approximation (dashed lines).} 
	\label{fig_slp_infl_linear_tau}
\end{figure}

The proposed diffusivity parameter depends on sediment characteristics (through $\tau_{ch}$ and $\varphi$)  and the bed shear stress, vanishing for large values of the latter (i.e. $\varepsilon_\beta \to 0$ for $\tau \to \infty$), in agreement with experimental observations. 
Fig. \ref{fig_slp_infl_linear_tau} compares the (exact) non-linear expression for $\Pi_\beta$ (eq. \ref{eq_ratio_qb_slp_ftau2}) against linear fits to $\Pi_\beta$ with line slopes given by \eqref{eq_diff_par_ftau}. 
Agreement of the linear fit with the original expression for $\Pi_\beta$ is very good for all negative bed slopes, mild adverse slopes, and large bed shear stress. 
Discrepancies increase for low $\tau$ and steep adverse slopes. 
The bed slope influence is well described by the linear approximation for $\tan \beta \lesssim 0.2 $ (in other words, $\beta \lesssim 11^{\circ}$ in keeping with steep slopes such as at gravel beaches) for all bed shear stresses herein considered. 
In fact, the linear approximation adheres well to the exact solution for $-\varphi \lesssim \beta \leq 0$ in all cases.  
The morphological diffusivity could have also been derived from the exact expression for $\Pi_\beta$ given by \eqref{eq_ratio_qb_slp_ftau2}; however, additional tests carried out by the authors indicated that the expression yielded is significantly more complicated and depends on more variables than \eqref{eq_diff_par_ftau}, with negligible quantitative improvement.

\subsection{Incorporation of morphological diffusion into a CM model} \label{sec:dif_exp_CM}

For this test, we define a CM model as the coupling between the 1D Non-linear Shallow Water Equations and the Exner equation via an empirical formula for bedload transport. 
We invoke a bedload formula of the form of Meyer-Peter \& M\"{u}ller's \cite{MPM1948}; namely: $\Phi = F_* (\theta -\theta_c)^{3/2}$; where $\Phi$ is the non-dimensional bedload transport; $F_*$ is a non-dimensional constant obtained from model calibration; and $\theta$ and $\theta_c$ are non-dimensional forms of bed shear stress and critical shear stress, respectively. 
Morphological diffusivity is incorporated into the CM model by use of \eqref{eq_qb_add_diff}. 
For comparison, the model by \cite{Bailard1981a} is also investigated. 
This model is of interest because \cite{Bailard1981a} proposed a formula for bedload on a sloping beach of the form of \eqref{eq_qb_add_diff}, from which an analytical expression for $\varepsilon_\beta$ may be deduced, provided that $\tan \beta \ll \tan \varphi$; namely: 
\begin{equation} \label{eq_dif_BI}
 \varepsilon_{B \& I} = \frac{1}{\tan \varphi} .
\end{equation}
 
We have derived the above expression by inserting the definition of bed slope influence, \eqref{eq_slope-infl-def}, in the formula by \cite{Bailard1981a}, and so $\varepsilon_{B \& I}$ is directly comparable to $\varepsilon_{\beta}$. 
Note that, although they have different roots, both $\varepsilon_{B \& I}$ and $\varepsilon_\beta$ predict an inverse proportionality with $\tan \varphi$. 
However, unlike $\varepsilon_{\beta}$, $\varepsilon_{B \& I}$ depends solely on the angle of repose and so is independent of bed shear stress (i.e. it does not vanish at large $\tau$). 
A similar expression to $\varepsilon_{B \& I}$ has been used by other authors in order to include the effect of the bed slope \cite[e.g.][]{Dodd2008}.

\begin{figure}
	\centering
	\includegraphics[height=6.0cm]{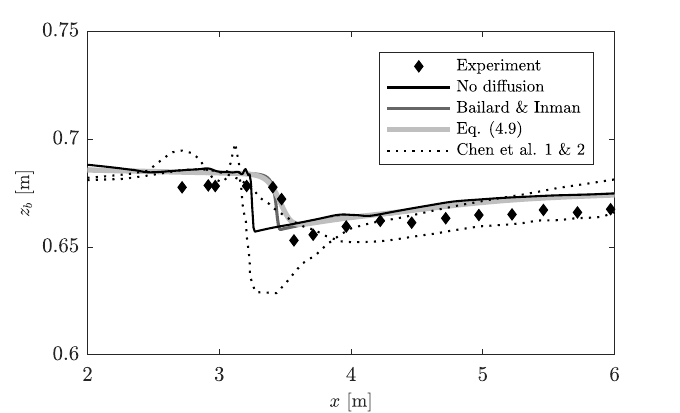}
	\caption{Final state of the mining pit bed profile predicted by a CM model including morphological diffusion derived from the present model ($\varepsilon_\beta$; eq. \ref{eq_diff_par_ftau}), from the work by Bailard \& Inman \cite{Bailard1981a} ($\varepsilon_{B \& I}$; eq. \ref{eq_dif_BI}), and for no diffusion at all. Comparison against experimental observations by \cite{Lee1993}. Predictions by \cite{Chen2010} shown in Fig. \ref{fig_q2l_pit} (Chen et al. 1 \& 2) are also included for reference.} 
	\label{fig_cmm_pit}
\end{figure}

We return to the mining pit investigated previously. 
The objectives are to test the effect that inclusion of morphological diffusion has on a CM model and the influence of bed shear stress on morphological diffusion (i.e. compare $\varepsilon_{\beta}$ vs $\varepsilon_{B \& I}$). 
Fig. \ref{fig_cmm_pit} illustrates this comparison. 
It is worth mentioning that in order to replicate faithfully the experimental set-up reported by \cite{Lee1993}, coefficient $F_*$ in the bedload equation takes a value of 2.3. 
This value is much lower than commonly used values proposed by \cite{MPM1948} (i.e. $F_* = 8$) and \cite{FernandezLuque1976} (i.e. $F_* = 5.7$). 
This implies some uncertainty in the reported set-up and appears to justify the use of a modified bed-update equation in the Q2L model (i.e. addition of $\eta_e$), as discussed in \S \ref{sec:model_validation}\ref{sec:experiment}. 

The CM model without morphological diffusion predicts the evolution of a very steep front and an unrealistic peak similar to one of the predictions by \cite{Chen2010}.  
Inclusion of morphological diffusion prevents development of the aforementioned peak, and both diffusivity parameters $\varepsilon_{\beta}$ and $\varepsilon_{B \& I}$ yield similar results, with the main difference between them being the slope of the predicted propagating front --- which is steeper in the latter case. 
Without more detailed experimental data it is difficult to reach a definite verdict regarding the proposed $\varepsilon_{\beta}$, but two points can be made with certainty: a) use of $\varepsilon_{\beta}$ yields realistic results whose agreement with experiments is at least of comparable quality to previous studies; and b) $\varepsilon_{\beta}$ and $\varepsilon_{B \& I}$ lead to similar results, although the former has the advantage of translating correctly hydrodynamic effects into the influence of bed slope on bedload (i.e. $\varepsilon_{\beta} \to 0$ for $\tau \to \infty$).

The same numerical scheme and set-up have been used for all predictions here presented. 
The CM model is discretised using second-order central differences in space, with fourth-order Runge-Kutta integration in time. 
All computations were undertaken on a uniform grid with $\Delta x = 0.01$ m and a time step $\Delta t = 0.01$ s. 
Discharge is imposed at both upstream and downstream boundaries; water depth is prescribed solely at the upstream boundary, and a transmissive condition is invoked at the downstream end; for the bed level, transmissive conditions are utilised in both boundaries. 
A ramp function is employed to bring the model from its initial condition of rest to the desired discharge, and when hydrodynamic steady state is reached, the sediment transport module is activated allowing the bed to evolve. 
However, it is important to highlight that even though numerical techniques have been widely utilised \cite[e.g.][]{Johnson2002} to prevent evolution of spurious high-frequency oscillations such as the peak predicted (in certain cases) upstream of the pit, the morphological diffusion here discussed is underpinned by phenomenological observations \cite[see e.g.][]{Moulton2014}. 
In other words, morphological diffusion is not merely a remedy for numerical instability but is also necessary for a realistic prediction of the morphodynamic model. 
Consider, for example, the case of the migrating hump studied by \cite{Hudson2003} within the framework of the present CM model. 
Under certain assumptions, the problem has an analytical solution by means of the method of characteristics, which, despite avoiding numerical oscillations, predicts a vertical wall ($S_b \to \infty$) in the migrating feature, which is clearly unrealistic because gravity will cause downslope motion of the sediment particles (morphological diffusion) when the local slope is sufficiently steep.
 
\begin{figure} 
	\begin{center}
		\begin{minipage}{0.9\linewidth}
			\subfigure[medium sand]{
				\resizebox*{0.4\linewidth}{!}{\includegraphics{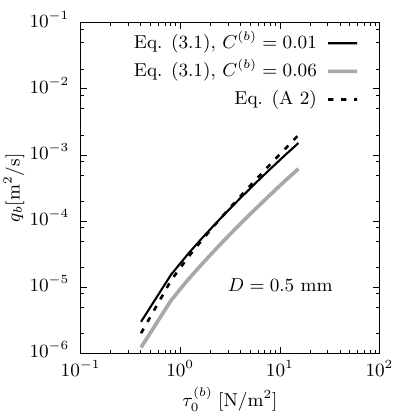}} } \hspace{5pt} 
			\subfigure[coarse sand]{
				\resizebox*{0.4\linewidth}{!}{\includegraphics{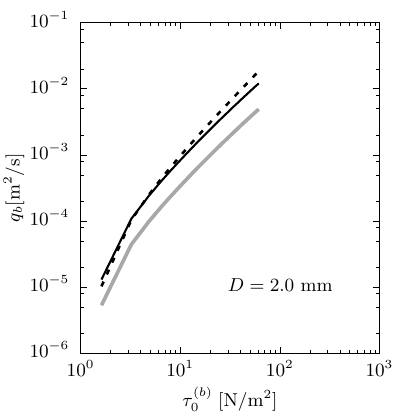}} }
			\caption{Comparison between bedload predicted by $q_b = f(u_{s 0})$ (eq. \ref{eq_qbsspexact}) and $q_b = f(u_0)$ (eq. \ref{eq_qb-mode1-ftau}) for two values of $C^{(b)}$ and for two different particle diameters.}
			\label{fig_appen}
		\end{minipage}
	\end{center}
\end{figure}
\section{Conclusions} \label{sec:conclusions}

A simplified 2-layer model has been introduced for the prediction of sediment transport rates and morphological evolution for bedload-dominated scenarios. 
The model differs from previous 2-layer models primarily through its treatment of the lower layer, which is here modelled as a thin layer of arbitrary but realistically small thickness. 
Thus, the inherent ambiguity in the definition of a near-bed transport layer is incorporated within the mathematical framework. 
Model results are found to be weakly dependent on the arbitrary selection of the lower layer thickness, within the range of values studied; a value of $h_0 \approx 10 D$, in agreement with previous works on bedload, is recommended. 
Although the model is devised specifically for bedload-dominated problems (such that $c_1 \to 0$), relaxation of the assumption of an upper layer consisting of pure water allows total load to be modelled to satisfactory accuracy (see Fig. \ref{fig_total_transport_1}) provided the concentration of suspended sediment is low. 
The Q2L model, successfully validated against empirical expressions for bedload, has then been compared against the experiment of a migrating mining pit by \cite{Lee1993}, with satisfactory agreement that rivals predictions by \cite{Chen2010}. 
An analytical expression has been derived for morphological diffusion (eq. \ref{eq_diff_par_ftau}) which permits easy modification of bedload empirical formulae, originally derived for nearly horizontal flumes, in order to render them applicable to steep stream-wise slopes, provided the bed slope angle, $| \beta | \lesssim 11^{\circ}$. 
The derived morphological diffusion is physically meaningful as it vanishes at large flow velocities, in agreement with previous experimental findings. 
Inclusion of the proposed morphological diffusivity in a CM model proves to enhance the latter by yielding better results, in comparison with the no-diffusion case, when applied to the mining pit experiment. 
It is thus shown that despite some simplifying assumptions, the proposed model and associated findings may improve the accuracy of morphodynamic models, especially in bedload-dominated environments where local bed slope may have an important influence on the bed evolution, such as in mountain rivers or steep gravel beaches. 
Current limitations of the Q2L model may be amended in the future by: considering non-uniform sediment size distributions, carrying out a parameter study for key variables and closure relationships, such as $C^{(b)}$, $C^{(i)}$, $\tau_b^{(b)}$, $\tau_{c \beta}$, $\tau_0^{(i)}$ and $\tau_0^{(b)}$, and revisiting certain assumptions like $u_{s 0} = u_0$ or $\rho_b = $ constant.\\

\enlargethispage{20pt}

\ethics{Not relevant to this work.}

\dataccess{This article does not include new experimental data.}

\aucontribute{SM and AGLB planned the research. SM developed and programmed the model, carried out the simulations and interpreted the results. Both the authors wrote the manuscript.}

\competing{We have no competing interests.}

\funding{SM was supported by the Mexican National Council for Science and Technology (Conacyt) through scholarship No. 310043 and the Mexican Ministry of Education (SEP).}

\ack{We thank the anonymous referees, whose insightful comments have led to a much improved paper. SM would like to thank The University of Edinburgh and Stanford University, where he was previously based and where much of this work was developed, and Joanna A. Zieli\'{n}ska for her help in reviewing some of the derivations.}

 
\appendix
\section{}\label{appA}
In order to study the relationship between $u_{s 0 }$ and $u_0$, an expression of the form $u_{s 0 } / u_* = \widehat{a} + \widehat{b} \ln D_* - \widehat{c} \, T_*^{-1/2} $ \cite{vRijn1984,Engelund1967} is employed, where $\widehat{a}$, $\widehat{b}$ and $\widehat{c}$ are tuning parameters; $D_* \equiv D [(s-1) g / \nu^2]^{1/3}$ is the non-dimensional particle diameter, with $s \equiv \rho_s / \rho_w$ and $\nu$ being the kinematic viscosity of water; $u_*$ is the bed friction velocity; and $T_* \equiv \tau_0^{(b)} / \tau_c$ is the transport stage. 
Considering that $\tau = \rho_w u_*^2 = \tau_0^{(b)} = \rho_0 C^{(b)} u_0^2$, the stream-wise bedload particle velocity for a given particle diameter can be written as:
\begin{equation} \label{eq_us0}
u_{s0} = \widehat{d} \left ( \frac{\rho_0 C^{(b)}}{\rho_w} \right )^{1/2} u_0 - \widehat{c} \left ( \frac{\tau_c}{\rho_w} \right )^{1/2} ,
\end{equation}
where $\widehat{d} = \widehat{a} + \widehat{b} \ln D_*$. 
Invoking equilibrium sediment transport considerations described in \S \ref{sec:model_validation}\ref{sec:bedload-analyt} to determine $c_0$, and using the substitution $u_0 = \sqrt{\tau_0^{(b)} / \left(\rho_0 C^{(b)}  \right) } $, we may write $q_b = h_0 c_0 u_{s 0}$ as:
\begin{equation} \label{eq_qbsspexact}
q_b = \frac{ \left ( \tau_0^{(b)} - \tau_c \right ) }{ (\rho_s - \rho_w) g \tan \varphi} \left [ \widehat{d} \left (\frac{\tau_0^{(b)}}{\rho_w}  \right )^{1/2} - \widehat{c} \left ( \frac{\tau_c}{\rho_w} \right )^{1/2} \right ] .
\end{equation}

Note that the above equation follows the form of bedload expressions proposed by \cite{Ashida1972} and \cite{Madsen1991}. 
It should be noted that, unlike \eqref{eq_qb-mode1-ftau}, the above equation does not depend on $C^{(b)}$ nor $h_0$; instead, calibration values of $\widehat{c}$ and $\widehat{d}$ are required. 
Based on numerical experiments within a Lagrangian framework, \cite{Maldonado2015} found values of $\widehat{d} = 8.328 + 1.328 \ln D_*$ and $\widehat{c} = 6.232$. 
Fig. \ref{fig_appen} compares the bedload rate predicted by \eqref{eq_qbsspexact}, using the aforementioned values of the coefficients, against the analytical solution derived in \S \ref{sec:model_validation}\ref{sec:bedload-analyt} (eq. \ref{eq_qb-mode1-ftau}). Two particle diameters are considered. 
For $C^{(b)} \approx 0.01$, both expressions yield very similar results. 
However, this is only the case for the values of calibration coefficients $\widehat{a}$, $\widehat{b}$ and $\widehat{c}$ considered here, which are particular to the ranges of particle diameter and flow velocity investigated by \cite{Maldonado2015}. 
Fig. \ref{fig_appen} confirms that the results and analysis presented throughout this paper for the Q2L model, based on the assumption $u_0 = u_{s0}$, appear valid from a practical viewpoint.

\section{} \label{AppB}
Consider an `active layer' of erodible bed material, of thickness, $z'_b$, and depth-variable density, $\rho'_b(z)$; the layer is located such that its upper face corresponds with the bed interface at $z_b$ (i.e. the layer is an upper sub-layer of $L_b$). 
The layer's average density, $\bar{\rho'_b}$, is estimated from the density evaluated at a point, $z_{\kappa}$, between $z=z_b - z'_b$ and $z=z_b$; i.e. $\bar{\rho'_b} \equiv \rho'_b(z=z_{(\kappa)})$, where $z_{(\kappa)} \equiv z_b - \kappa z'_b$.  
Defining the mass exchange through the bed surface interface as $\bar{\rho'_b} e^{(b)}$, mass conservation applied to the layer yields:
\begin{align*}
\frac{\mathrm{d} (\bar{\rho'_b} z'_b) }{\mathrm{d} t} &= \bar{\rho'_b} \frac{\partial z'_b}{\partial t} + z'_b  \left. \left [ \frac{\partial \bar{\rho'_b} }{\partial z} \frac{\partial z}{\partial t} \right ] \right |_{z=z_{(\kappa)}} = -\bar{\rho'_b} e^{(b)} \\
&= \bar{\rho'_b} \frac{\partial z'_b}{\partial t} + \left. \frac{\partial \bar{\rho'_b} }{\partial z} \right |_{z=z_{(\kappa)}} \left ( \frac{\partial z_b}{\partial t} - \kappa \frac{\partial z'_b}{\partial t} \right ) = -\bar{\rho'_b} e^{(b)}  ,
\end{align*}
where $\partial \bar{\rho'_b} / \partial t = 0$ has been assumed. 
Further noting that $\partial z_b / \partial t = \partial z'_b / \partial t$, it follows that: 
\begin{equation*}
\frac{\partial z_b}{\partial t} = \left ( - \frac{ \bar{\rho'_b} }{\bar{\rho'_b} + z'_b (1-\kappa) \left. \frac{\partial \bar{\rho'_b}}{\partial z} \right |_{z=z_{(\kappa)}} } \right ) e^{(b)} = -\eta_e e^{(b)} ,
\end{equation*}
from which it is clear that $\eta_e$ relates to the vertical profile of the bed average density, $\partial \bar{\rho'_b} / \partial z$, over the thickness $z'_b$ of an active erodible layer, which is expected to be different from zero in real experiments, such as the one considered here. 

\emph{Remark.} If $\rho'_b(z)$ varies linearly over $z'_b$, then $\kappa = 1/2$ and  
\begin{equation*}
\left. \frac{\partial \bar{\rho'_b} }{\partial z} \right |_{z=z_{(1/2)}} = \frac{\rho'_b(z=z_b) - \rho'_b(z=z_b - z'_b)}{z'_b} .
\end{equation*}

\section{Notation list}
$c_k(x,t) = $ sediment concentration of layer $L_k$ ($k=0,1$);\\ 
$c_b=$ bed sediment concentration;\\ 
$c_{0 \, \textup{mx}}=$ maximum allowed value for $c_0$ (saturation concentration for $L_0$); \\
$(C^{(b)}, C^{(i)}) = $ Q2L model calibration parameters; \\
$D=$ sediment particle diameter; \\
$e^{(b)}=$ net water-sediment volumetric exchange between $L_b$ and $L_0$ (the erosion rate); \\
$g=$ gravitational acceleration; \\
$h_k=$ thickness (depth) of layer $L_k$ ($k=0,1$); \\
$i^n=$ net water-sediment mass exchange between layers through interface $n$ ($n=(b),(i)$); \\
$i_s^n=$ net sediment mass exchange between layers through interface $n$ ($n=(b),(i)$); \\
$j^n=$ net horizontal momentum exchange between layers through interface $n$ ($n=(b),(i)$); \\
$L_k (k=b,0,1)=$ bed, lower and upper layer, respectively; \\
$q_b=$ volumetric bedload transport rate per unit width; \\
$q_T = $ total (bedload + suspended) sediment transport rate per unit width; \\
$\bar{u} = $ whole-depth-averaged streamwise velocity; \\
$u_k(x,t) = $ parallel-to-bed streamwise velocity of layer $L_k$ ($k=b,0,1$); \\
$u_{sk}=$ streamwise sediment particle velocity within layer $L_k$ ($k=0,1$); \\
$S_b = \tan \beta = \partial z_b / \partial x =$ bed slope; \\
$t=$ time; \\
$(x,z)=$ Cartesian frame of reference with streamwise and vertical coordinates, respectively; \\
$z_b=$ bed level with respect to a datum; \\
$\beta =$ bed slope angle; \\
$\varepsilon_\beta =$ morphological diffusivity parameter; \\
$\varepsilon_{B \& I}=$ morphological diffusivity derived from the work by \cite{Bailard1981a}; \\
$\eta_e =$ additional Q2L model tuning parameter related to non-uniform bed material and packing fraction; \\
$\Pi_{\beta} =$ bed slope influence as defined in \eqref{eq_slope-infl-def}; \\
$\rho_k=$ bulk density of layer $L_k$ ($k=b,0,1$); \\
$\rho_s=$ density of sediment; \\
$\rho_w=$ density of water; \\
$\tau_c = $ critical bed shear stress; \\
$\tau =$ bed shear stress; \\
$\tau_0^{(b)} =$ shear stress exerted by the fluid on the bed surface; \\
$\tau_b^{(b)} =$ bed resistance to erosion; \\
$(\tau_0^{(i)}, \tau_1^{(i)}) =$ shear stress at the bottom and top of interface $(i)$, respectively; \\
$\varphi =$ angle of repose; and\\
subscripts $\beta$ and $h$ denote sloping and horizontal bed, respectively.

\bibliographystyle{RS}
\bibliography{Q2LpaperRSOS}

\end{document}